\documentstyle[12pt]{article}
\newcommand{\newc}{\newcommand}
\newc{\beq}{\begin{equation}}
\newc{\eeq}{\end{equation}}
\newc{\bea}{\begin{array}}
\newc{\eea}{\end{array}}
\newc{\pd}{\partial}
\newc{\Psibar}{\overline\Psi}
\newc{\qbar}{\overline q}
\newc{\w}{{\bf w}}
\newc{\bfn}{{\mathbf\nabla}}
\newc{\bfg}{{\mathbf\gamma}}
\newc{\E}{{\mathbf{E}}}
\newc{\bp}{{\bf p}}
\newc{\la}{{\cal L}}
\newc{\ti}{{\times}}
\newc{\bA}{{\bf A}}
\newc{\ri}{{\rm i}}
\newc{\bL}{{\bf L}}
\newc{\bS}{{\bf S}}
\newc{\bB}{{\mathbf B}}
\newc{\bfx}{{\bf x}}
\newc{\bfV}{{\bf V}}
\newc{\bu}{{\bf u}}
\newc{\bM}{{\bf M}}
\newc{\bV}{{\bf V}}
\newc{\bv}{{\bf v}}
\newc{\bx}{{\bf x}}
\newc{\bD}{{\bf D}}
\newc{\bH}{{\bf H}}
\newc{\br}{{\bf r}}
\newc{\ve}{{\varepsilon}}
\newc{\bb}{{\bf b}}
\newc{\bc}{{\bf c}}
\newc{\bj}{{\bf j}}
\newc{\bd}{{\bf d}}
\newc{\bE}{{\mathbf{E}}}
\newc{\tla}{{\tilde{\cal L}}}
\newc{\ho}{\hookrightarrow }
\newc{\bP}{{\bf P}}
\newc{\piv}{{\partial_4}}
\newc{\pv}{{\partial_5}}
\newc{\bze}{{\mathbf 0}}

\newc{\sig}{{\mathbf\sigma}}
\newc{\bpi}{{\mathbf\pi}}
\newc{\eg}{{\rm e.g.\ }}
\newc{\ie}{{\rm i.e.\ }}
\newc{\etal}{{\rm et al\ }}
\def\JPA#1#2#3#4{#2 #1 {\it J. Phys. A: Math. Gen.} {\bf #3} #4}
\def\BJPS#1#2#3#4{#2 #1 {\it Brit. J. Phil. Sci.} {\bf #3} #4}
\def\PS#1#2#3#4{#2 #1 {\it Phil. Sci.} {\bf #3} #4}

\def\NGWG#1#2#3#4{#2 #1 {\it Nachr. Ges. Wiss. Gottingen} {\bf #3} #4}

\def\MAG#1#2#3#4{#2 #1 {\it Math. Annalen} {\bf #3} #4}
\def\AP#1#2#3#4{#2 #1 {\it Ann. Phys. (NY)} {\bf #3} #4}
\def\CMP#1#2#3#4{#2 #1 {\it Comm. Math. Phys.} {\bf #3} #4}
\def\CRASP#1#2#3#4{#2 #1 {\it Comptes Rendus de l'Acad\'emie des Sciences de Paris}
 {\bf #3} #4}
\def\PLA#1#2#3#4{#2 #1 {\it Phys. Lett. A} {\bf #3} #4}
\def\EE#1#2#3#4{#2 #1 {\it Electrical Engineering (Archiv fur Elektrotechnik)}
 {\bf #3} #4}
\def\APG#1#2#3#4{#2 #1 {\it Annalen Phys.} {\bf #3} #4}
\def\AHES#1#2#3#4{#2 #1 {\it Archive for History of Exact Sciences}
 {\bf #3} #4}

\def\SHPMP#1#2#3#4{#2 #1 {\it Studies in the History and Philosophy of Modern
 Physics} {\bf #3} #4}
\def\EPL#1#2#3#4{#2 #1 {\it Europhys. Lett.} {\bf #3} #4}
\def\AFLB#1#2#3#4{#2 #1 {\it Ann. Fondation Louis de Broglie} {\bf #3} #4}

\def\RPJ#1#2#3#4{#2 #1 {\it Russian Phys. J.} {\bf #3} #4}
\def\EJP#1#2#3#4{#2 #1 {\it Eur. J. Phys.} {\bf #3} #4}
\def\PTRSA#1#2#3#4{#2 #1 {\it Phil. Trans. R. Soc.} {\bf A #3} #4}

\def\NCB#1#2#3#4{#2 #1 {\it Nuov. Cim.} {\bf B #3} #4}
\def\RSI#1#2#3#4{#2 #1 {\it Rev. Sc. Instrum.} {\bf #3} #4}
\def\CENTAURUS#1#2#3#4{#2 #1 {\it Centaurus} {\bf #3} #4}

\def\PTRSL#1#2#3#4{#2 #1 {\it Phil. Trans. Roy. Soc. London} {\bf #3} #4}

\def\AJP#1#2#3#4{#2 #1 {\it Am. J. Phys.} {\bf #3} #4}

\catcode`@=11
\long
\def\@caption#1[#2]#3{\par\addcontentsline{\csname
  ext@#1\endcsname}{#1}{\protect\numberline{\csname
  the#1\endcsname}{\ignorespaces #2}}\begingroup
    \small
    \@parboxrestore
    \@makecaption{\csname fnum@#1\endcsname}{\ignorespaces #3}\par
  \endgroup}
\catcode`@=12

\begin{document}

\begin{titlepage}
\vskip 2cm
\begin{center}
{\Large\bf On some applications of Galilean electrodynamics of moving bodies
\footnote{E-mail: {\tt montigny@phys.ualberta.ca,}
 {\tt Germain.Rousseaux@inln.cnrs.fr}\hfill}}
\vskip 3cm
{\bf
M. de Montigny$^{a,b}$ and G. Rousseaux$^{c}$ \\}
\vskip 5pt
{\sl $^a$Campus Saint-Jean, University of Alberta \\
 8406 - 91 Street \\
 Edmonton, Alberta, Canada T6C 4G9\\}
\vskip 2pt
{\sl $^b$Theoretical Physics Institute, University of Alberta\\
 Edmonton, Alberta, Canada T6G 2J1\\}
\vskip 2pt
{\sl $^c$Universit\'e de Nice Sophia-Antipolis, Institut Non-Lin\'eaire de Nice,\\
 INLN-UMR 6618 CNRS-UNICE, 1361 route des Lucioles,\\
 06560 Valbonne, France\\}
\vskip 2pt

\end{center}
\vskip .5cm
\rm
\begin{abstract}
We discuss the seminal article in which
 Le Bellac and L\'evy-Leblond have identified two Galilean
 limits of electromagnetism \cite{bellac}, and its modern
 implications.  Recent works have shed a new light on the choice
 of gauge conditions in classical electromagnetism. We discuss various applications and
 experiments, such as in quantum mechanics, superconductivity,
 electrodynamics of continuous media, etc. Much of the current technology, where
 waves are not taken into account, is actually based
 on Galilean electromagnetism.
\end{abstract}
\end{titlepage}

\setcounter{footnote}{0} \setcounter{page}{1} \setcounter{section}{0} %
\setcounter{subsection}{0} \setcounter{subsubsection}{0}

\section{Introduction}

The purpose of this article is to emphasize the relevance of `Galilean
 electromagnetism', recognized in 1973 by Michel Le Bellac and Jean-Marc
 L\'evy-Leblond (LBLL). They observed that there exist not only one,
 but {\em two} well-defined Galilean limits of electromagnetism:
 the so-called `magnetic' and `electric' limits \cite{bellac}.
 Hereafter, `Galilean' means that the theory satisfies the principle of relativity in
 its Galilean form (also referred to as the sometimes misleading term `non-relativistic').

Our purpose is {\em not} to argue that these limits
 should be seen as alternatives to Lorentz-covariant electrodynamics. We wish to
 point out that some physical phenomena, often described with special relativity,
 can be explained by properly defined Galilean limits.
 In other words, such phenomena could have been understood without recourse
 to special relativity, had the Galilean limits of electrodynamics been
 correctly defined in the first place. Therefore,
 our general purpose is twofold: first, that one must be careful when investigating
 alleged `non-relativistic' limits, and second, that well-defined Galilei-covariant
 theories might allow one to describe more physical phenomena than usually
 believed. The later point means that some concepts, which are thought to be
 `purely relativistic', can actually be understood within
 the realm of Galilean physics. A dramatic such example is
 the concept of spin \cite{levyleblond1967}.

We have summarized and discussed various approaches to Galilean
 electromagnetism in a recent article \cite{ejp}. Since then, we have learned
 the existence of several studies emphasing the applications of quasistatic
 regimes both in research like, for example, in micro-electronics \cite{Dirks},
 bio-systems engineering, medical engineering, electromagnetic computations
 \cite{Ursula} and teaching \cite{Larsson}.

Now let us briefly review Galilean electromagnetism
 and set up the main equations that we are
 going to utilize later on. A {\em Lorentz transformation} acts on
 space-time coordinates as follows:
\beq\bea{l}
\bx'=\bx-\gamma\bv t+(\gamma-1)\frac{\bv (\bv\cdot\bx)}{\bv^2},\\
t'=\gamma\left (t-\frac{\bv\cdot\bx}{c^2}\right),
\eea\label{xtlorentz}\eeq
where $\bv$ is the relative velocity and $\gamma=\frac 1{\sqrt{1-(v/c)^2}}$.
 When $v<<c$, this reduces to
 a {\em Galilean transformation} of space-time:
\beq\bea{l}
\bx'=\bx-\bv t,\\
t'=t.
\eea\label{xtgalilei}\eeq
Since Galilean kinematics involves the time-like condition
\beq
c\Delta t>>\Delta x,\label{timelikecondition}
\eeq
 there is no other possible limit than the one given in Eq. (\ref{xtgalilei}).
 As we shall see below, this is not the same for transformations of
 electric and magnetic fields.

Under a Lorentz transformation, Eq. (\ref{xtlorentz}), the electric and magnetic fields
 in vacuum transform as
\beq\bea{l}
\bE'=\gamma(\bE+\bv\times\bB)+(1-\gamma)\frac{\bv (\bv\cdot\bE)}{\bv^2},\\
\bB'=\gamma(\bB-\frac 1{c^2}\bv\times\bE)+(1-\gamma)\frac{\bv (\bv\cdot\bB)}{\bv^2}.
\eea\label{fieldslorentz}\eeq
 If we take the limit $v/c\rightarrow 0$, we find
\beq\bea{l}
\bE'=\bE+\bv\times\bB,\\
\bB'=\bB.
\eea\label{premaglim}\eeq
As we will see shortly, this is a legitimate limit called the
 `magnetic limit' of electromagnetism.
 (One might be tempted to consider the limit $\gamma\rightarrow 1$ which
 leads to
\beq\bea{l}
\bE'=\bE+\bv\times\bB,\\
\bB'=\bB-\frac 1{c^2}\bv\times\bE.
\eea\label{wrongfieldtrans}\eeq
However, it is not a valid transformation; in particular,
 it does not even satisfy the group composition law
 \cite{bellac}.) However, Eq. (\ref{fieldslorentz}) allows
 us to obtain, in addition to Eq. (\ref{premaglim}), another perfectly
 well-defined Galilean limit. In order to do so, one must compare the modules
 of the electric field $E$ and the magnetic field $cB$, in analogy with
 Eq. (\ref{timelikecondition}). For large magnetic fields,
 Eq. (\ref{fieldslorentz}) reduces to the so-called
 {\it magnetic limit} of electromagnetism:
\beq\bea{l}
\bE_m'=\bE_m+\bv\times\bB_m,\qquad\qquad E_m<<cB_m,\\
\bB_m'=\bB_m.
\eea\label{fieldsmagnetic}\eeq
The alternative, for which the electric field dominates, leads to the
 {\it electric limit}:
\beq\bea{l}
\bE_e'=\bE_e,\qquad\qquad E_e>>cB_e,\\
\bB_e'=\bB_e-\frac 1{c^2}\bv\times\bE_e.
\eea\label{fieldselectric}\eeq
Indeed, the approximations $E_e/c>>B_e$ and $v<<c$ together imply that
 $E_e/v>>E_e/c>>B_e$, so that we take $E_e>>vB_e$ in Eq. (\ref{fieldslorentz}).

Since we will emphasize the use of scalar and vector potentials $(V,\bA)$, let us
 consider their transformation properties. Under a Lorentz transformation,
 Eq. (\ref{xtlorentz}), they become
\beq\bea{l}
\bA'=\bA-\frac{\gamma\bv V}{c^2}+(\gamma-1)\frac{\bv (\bv\cdot\bA)}{\bv^2},\\
V'=\gamma\left (V-{\bv\cdot\bA}\right).
\eea\label{AVlorentz}\eeq
When $v<<c$ and $A<<cV$, this reduces to the {\em electric limit}
 of potential transformations:
\beq\bea{l}
\bA_e'=\bA_e-\frac{\bv V_e}{c^2},\\
V_e'=V_e.
\eea\label{elAVgalilei}\eeq
 The electric and magnetic fields are expressed in terms of the potentials
 as follows
\beq
{\bf E}_e= -\nabla V_e,\qquad\quad {\bf B}_e=\nabla\times A_e.
\label{elFvsA}\eeq
 Whereas there exists only one possible condition, Eq. (\ref{timelikecondition}),
 for the space-time manifold, here we find a second limit, obtained
 by $v<<c$ and $A>>cV$, such that Eq. (\ref{AVlorentz}) reduces to the
 {\em magnetic limit} of potential transformations:
\beq\bea{l}
\bA_m'=\bA_m,\\
V_m'=V_m-\bv\cdot\bA_m.
\eea\label{magAVgalilei}\eeq
 In this limit, the electromagnetic field components are given
 by
\beq
{\bf E}_m=-\nabla V_m-\partial_t {\bf A}_m,
\qquad\quad {\bf B}_m=\nabla\times A_m.
\label{magFvsA}\eeq

Finally, let us recall the two Galilean limits of the Maxwell
 equations. Their relativistic form is written as
\beq\bea{rcll}
\bfn\ti\bE&=&-\pd_t\bB,\quad & {\rm Faraday},\\
\bfn\cdot\bB&=&0,\quad & {\rm Thomson},\\
\bfn\ti\bB&=&\mu_0\bj+\frac 1{c^2} \pd_t\bE,\quad & {\rm Ampere},\\
\bfn\cdot\bE&=&\frac 1{\epsilon_0}{\rho},\quad & {\rm Gauss},
\eea\label{maxwell}\eeq
 The existence of two Galilean limits is not so obvious
 if one naively takes the limit $c\rightarrow\infty$.
 LBLL have found in Ref. \cite{bellac} that, in the electric limit,
 the Maxwell equations reduce to:
\beq\bea{l}
\bfn\ti\bE_e=\bze ,\\
\bfn\cdot\bB_e=0 ,\\
\bfn\ti\bB_e-\frac 1{c^2} \pd_t\bE_e=\mu_0\bj_e ,\\
\bfn\cdot\bE_e=\frac 1{\epsilon_0}{\rho_e}.\eea
\label{maxel}\eeq
Clearly, the main difference with the relativistic Maxwell equations is that
 here the electric field has zero curl in Faraday's law.
In the magnetic limit, the Maxwell equations become
\beq\bea{l}
\bfn\ti\bE_m=-\pd_t\bB_m ,\\
\bfn\cdot\bB_m=0,\\
\bfn\ti\bB_m=\mu_0\bj_m ,\\
\bfn\cdot\bE_m=\frac 1{\epsilon_0}\rho_m .\eea\label{maxmag}
\eeq
The displacement current term is absent in Amp\`ere's law.

Hereafter, we illustrate some applications of the Galilean electrodynamics of moving bodies.
 In the next section, we reexamine the gauge conditions and their compatibility with
 Lorentz and  Galilean covariance. Then we comment briefly on the connection between the
 two limits and the Faraday tensor (and its dual). In Section 4, we discuss Feynman's proof
 of (the magnetic limit of) the Maxwell equations, and Section 5 contains a few comments
 about superconductivity seen as a magnetic limit, and gauge potentials. In Sections
 6 and 7, we question
 our current understanding of the electrodynamics of moving bodies by examining the
 Trouton-Noble experiment in a Galilean context as well the introductory example used by Einstein
 in his famous work on special relativity.
 We conclude with some comments on the intrinsic use by Maxwell of both limits,
 one century before LBLL.

%

%  GAUGE CONDITIONS

\section{Gauge conditions and Galilean electromagnetism}

Hereafter, we use the Riemann-Lorenz formulation of classical electromagnetism
 (i.e. in terms of scalar and vector potentials instead of fields \cite{Alfred}) to
 describe the two Galilean limits. Let us recall how the electric and magnetic
 limits may be retrieved in this formulation by a careful
 consideration of orders of magnitude \cite{europhys,ejp}.
 It is quite natural to define the following dimensionless parameters:
\beq
\ve\equiv\frac L{cT}\qquad {\rm and}\qquad \xi\equiv\frac j{c\rho},
\label{epsilonxi}\eeq
where $L$, $T$, $j$ and $\rho$ represent the orders of magnitude of
 length, time, current density, and charge density, respectively.

The equations of classical electromagnetism, written in terms of potentials, are cast
 into the following form \cite{Alfred}:
\beq\bea{l}
\nabla^2V-\frac 1{c^2}\frac{\partial^2 V}{\partial t^2}=-\frac\rho{\epsilon_0},\qquad
 {\rm Riemann\ equations},\\
\nabla^2\bA-\frac 1{c^2}\frac{\partial^2 \bA}{\partial t^2}=-\mu_0\bj,\eea\label{riemann}
\eeq
\beq
\bfn\cdot\bA+\frac 1{c^2}\frac{\partial V}{\partial t}=0,\qquad {\rm Lorenz\ equation},
\label{lorentz}\eeq
\beq
\frac {\rm d}{{\rm d}t}(m\bv+q\bA)=-q\nabla _{\bA} (V-\bv\cdot\bA),\qquad {\rm Lorentz\ force}.
\eeq
The quasistatic approximation, $\ve<<1$, of Eq. (\ref{riemann}) leads to
\beq
\nabla^2V\simeq -\frac\rho{\epsilon_0}\qquad\qquad {\rm and}\qquad\qquad
 \nabla^2 {\bf A}\simeq -\mu_0 {\bf j},
\label{potentialvscurrent}\eeq
from which we can define a further dimensionless ratio,
$\frac {cA}V\simeq\frac j{\rho c}$, so that
\beq
\frac {cA}V\simeq \xi.
\label{potxi}\eeq
This echoes LBLL's prescription for the fields \cite{bellac}:
 in the magnetic limit, the spacelike quantity $cA$ is dominant, whereas in
 the electric limit, it is the timelike quantity $V$ that dominates.

The definition ${\bf E}=-\partial_t {\bf A}-\nabla V$
 of the electric field takes different forms in the Galilean limits,
 depending on the order of magnitude of each term,
 because the Galilean transformations for the potentials
 differ for the electric and the magnetic limits \cite{bellac}.
 Let us evaluate the order of magnitude of the
 ratio between its two terms:
\[
\frac{\partial_t {\bf A}}{\nabla V}\simeq
 \frac{\frac{A}{T}}{\frac{V}{L}}
 \simeq \frac{L}{cT}\frac{cA}{V} \simeq \ve \xi.
\]
In the magnetic limit, for which $\xi>>1$, this equation
 leads to Eq. (\ref{magFvsA}). By computing the curl, we find
 $\partial_t {\bf B}_m=-\nabla \times {\bf E}_m$.
 Likewise, in the electric limit, for which  $\xi<<1$, we can
 neglect $\partial_t\bA$, so that we obtain
 Eq. (\ref{elFvsA}). The curl of this expression leads to
 $\nabla \times {\bf E}_e\simeq {\bf 0}$.

The choice of gauge conditions allows one to retrieve the two sets of Galilean
 Maxwell equations in terms of fields, as stated by LBLL \cite{bellac}.
 Moreover, as we now proceed to show, the gauge
 conditions are closely related to the nature of kinematic transformations.
 In the magnetic limit, the condition $\xi>>1$ leads to the Coulomb gauge condition:
 $\bfn\cdot\bA_m=0$.
 From the definition of $\bB_m$ together with the identity
\beq
\nabla \times(\nabla \times {\bf A})=\nabla(\nabla
 \cdot{\bf A})-\nabla ^2 {\bf A},
\label{identity}\eeq
 we see that
\[
\nabla \times{\bf B}_m=\nabla\times (\nabla \times {\bf A}_m)=
 \nabla(\overbrace{\nabla\cdot\bA_m}^0)-\nabla^2\bA_m=\mu_0\bj_m.
\]
The last term follows from Eq. (\ref{potentialvscurrent}).
 We point out that the displacement current term is missing.
 The divergence of the electric field, in the magnetic limit, gives
\[
\nabla\cdot\bE_m=\nabla\cdot (-\partial_t {\bf A}_m-\nabla V_m)
 =-\partial_t(\overbrace{\nabla\cdot\bA_m}^0)-\nabla^2V_m=
\frac{\rho_m}{\epsilon_0},
\]
where we have utilized Eq. (\ref{identity}). This is the second
 inhomogeneous equation, in the last line of Eq. (\ref{maxmag}).
 In the electric limit, the condition $\xi<<1$ leads similarly to
 the Lorenz condition. Proceeding as in
 the magnetic limit, we begin with the curl of $\bB_e$:
\[
\nabla \times{\bf B}_e=\nabla\times (\nabla \times {\bf A}_e)=
 \nabla(\overbrace{\nabla\cdot\bA_e}^{-(\partial_t V_e)/c^2})-\nabla^2\bA_e
 =\frac 1{c^2}\partial_t\bE_e+\mu_0\bj_e.
\]
>From the divergence of $\bE_e$, we find
\[
\nabla\cdot\bE_e=\nabla\cdot (-\nabla V_e)=-\nabla^2 V_e=
\frac{\rho_e}{\epsilon _{0}},
\]
where we have used (\ref{potentialvscurrent}).

To summarize the preceding discussion, we point out forcefully that the choice of
 a gauge condition is dictated by the relativistic versus Galilean nature of the problem.
 The role of potentials and gauge conditions in quasistatic regimes was
 pointed out only recently by Dirks \cite{Dirks} and Larsson \cite{Larsson},
 although the problem was not handled correctly, as we have done with the Riemann-Lorenz formulation. Indeed, both of them use the (erroneous) equation :
 \beq
 \nabla^2 {\bf A} -\frac 1{c^2}\frac{\partial^2 \bA}{\partial t^2}= -\mu_0 {\bf j}+\frac{1}{c^2}\nabla \frac{\partial V}{\partial t},
\label{potentialvscurrent}\eeq
obtained by plunging the (magnetic Galilean-covariant: $\xi>>1$) Coulomb gauge condition into the full set of (Lorentz-covariant:$\xi \simeq 1$) Maxwell equations in terms of the fields without remarking that the temporal terms are negligible with respect to the spatial terms as $\epsilon <<1$...
 
 The Lorenz gauge condition is compatible with
 the relativistic context as well as the electric Galilean limit.
 The Coulomb gauge condition, however, is compatible with the Galilean
 magnetic limit only because it is not covariant with respect to neither
 the Lorentz transformations nor the Galilean electric transformations.
 We refer the interested reader to a discussion of the physical meaning
 that one can ascribe to the various `gauge conditions' \cite{aflb}.

Galilean electromagnetism sheds a new light on the pre-relativity era.
 Indeed, a careful reading of James Clerk Maxwell's famous
 {\it Treatise on Electricity and Magnetism} reveals that he was actually working
 with the electric limit in the discussion of dielectric materials in his
 first volume (see Chapters II to V) \cite{Maxwell}.
 Likewise, in his treatment of ohmic conductors and induced magnetic fields, in his second
 volume, the magnetic limit was employed implicitly, except in the chapters on the
 theory of light propagation, where he introduced `by hand' the displacement current
 term into the magnetic limit equations in order to demonstrate that light
 is a transverse electromagnetic wave \cite{Maxwell}.  But,
 as we have seen in the particular case of the electric limit (and it is also valid in relativity),
 the displacement current follows from choosing the Lorenz gauge, and Maxwell
 (wrongfully) kept the Coulomb gauge within the relativistic context for
 the fields (for more details, see the forthcoming \cite{Pierseaux}). This problem prompted
 Hertz and Heaviside to relinquish potentials and to rather
 cast the Maxwell equations in terms of fields. Following
 Hertz's approach, Einstein subsequently expressed the
 Maxwell equations in terms of fields (i.e. in the
 Heaviside-Hertz formulation), whereas Henri Poincar\'e wrote the
 Maxwell equations in terms of the potentials
 (i.e. the Riemann-Lorenz formulation) by adopting the
 Lorenz condition in a relativistic context \cite{Pierseaux}.

\section{The Faraday tensor and its dual}

In special relativity, it is well known that the Faraday tensor:
\[
F_{\mu\nu}\equiv \partial_\mu A_\nu-\partial_\nu A_\mu,
 \qquad\quad \mu,\nu=0,1,2,3,
\]
 and its dual:
\[
^*F_{\mu\nu}=\frac 12\epsilon^{\mu\nu\rho\sigma}F_{\rho\sigma},
\]
 do have the same physical meaning. This is not the case
 within Galilean electromagnetism. As pointed out in Earman's book,
 and recently discussed by Rynasiewicz, the Galilean
 tranformations of the Faraday tensor and its dual tensor lead to the
 electric or the magnetic limit, respectively \cite{Earman}.
 The effect of the duality operation amounts to exchanging
 $\bE$ and $\bB$:
 \[
{\bf{E}}\rightarrow c{\bf{B}}\quad {\rm and}\quad
 {\bf{B}}\rightarrow -{\bf{E}}/c.
\]
 One recovers the magnetic and electric limits,
 Eqs. (\ref{fieldsmagnetic}) and (\ref{fieldselectric}),
 by applying the duality transformations
 directly to the electric transformations of the fields
 in order to get the magnetic transformations, and vice versa.

Earman also noted \cite{Earman} that the field transformations of
 the magnetic limit are obtained when $\bE$ and $\bB$ are expressed
 in terms of `covariant', or
 $\left (\begin{array}{cc} 0\\ 2\end{array}\right )$, tensor $F_{\mu\nu}$,
 whereas the electric limit is obtained when the fields transformations
 are calculated by using the `contravariant', or
 $\left (\begin{array}{cc} 2\\ 0\end{array}\right )$, tensor $F^{\mu\nu}$.
 Let us illustrate it briefly, with
\[
A^\mu=\left (\frac Vc,\bA\right ),\qquad A_\mu=\left (\frac Vc,-\bA\right ),
\]
as well as
\[
\partial^\mu=\left (\frac 1c\partial_t,-\nabla\right ),\qquad
 \partial_\mu=\left (\frac 1c\partial_t,\nabla\right ).
\]

The magnetic limit follows from  the relation
\[
F'_{\mu\nu} = \Lambda_\mu^{\ \rho} \Lambda_\nu^{\ \sigma}\; F_{\rho\sigma},
\]
where the Galilean transformation matrix $\Lambda_\mu^{\ \nu}$
 is defined by the four-gradient transformation,
 $\partial'_\mu=\Lambda_\mu^{\ \nu}\partial_\nu$,
 so that
\[
\Lambda_\mu^{\ \nu}=\left (
\begin{array}{cccc}
1 & \frac{v_x}c & \frac{v_y}c & \frac{v_z}c \\
0 & 1 & 0 & 0 \\
0 & 0 & 1 & 0 \\
0 & 0 & 0 & 1 \end{array}\right ).
\]
The index $\mu$ denotes the line of each entry.
We find, for example,
\[
\begin{array}{rcl}
 \frac{E'_x}c=F'_{01}&=& \Lambda_0^{\ \mu} \Lambda_1^{\ \nu}\; F_{\mu\nu},\\
 &=& \Lambda_0^{\ 0}F_{01} + \Lambda_0^{\ 2}F_{21} +
 \Lambda_0^{\ 3}F_{31},\\
 &=& \frac 1c (E_x+v_yB_z-v_zB_y),
\end{array}
\]
and
\[
-B'_z=F'_{12}=\Lambda_1^{\ \mu}\Lambda_2^{\ \nu}F_{\mu\nu}=-B_z,
\]
which is Eq. (\ref{fieldsmagnetic}).

The electric limit transformations follows from
\[
F'^{\mu\nu} = \Lambda^\mu_{\ \rho} \Lambda^\nu_{\ \sigma}\; F^{\rho\sigma}.
\]
The transformation matrix $\Lambda^\mu_{\ \nu}$
 is now defined by the coordinate transformation,
 $x^\mu=\Lambda^\mu_{\ \nu}\; x^\nu$, with
 $x^\mu=(ct,x,y,z)$, so that
\[
\Lambda^\mu_{\ \nu}=\left (
\begin{array}{cccc}
1 & 0 & 0 & 0 \\
-\frac{v_x}c & 1 & 0 & 0 \\
-\frac{v_y}c  & 0 & 1 & 0 \\
-\frac{v_z}c  & 0 & 0 & 1 \end{array}\right ).
\]
Again, the first index $\mu$ denotes the matrix line.
For instance, we compute
\[
-\frac{E'_x}c=F'^{01}=\Lambda^0_{\ 0}\Lambda^1_{\ \nu}F^{0\nu}=-\frac{E_x}c,
\]
and
\[
\begin{array}{rcl}
 {B'_z}=-F'^{12}&=&-\Lambda^1_{\ \mu} \Lambda^2_{\ \nu}\; F^{\mu\nu},\\
 &=& -\Lambda^1_{\ 0}F^{02} - \Lambda^2_{\ 0}F^{10} - F^{12},\\
 &=& B_z-\frac 1{c^2}(v_xE_y-v_yE_x),
\end{array}
\]
which is Eq. (\ref{fieldselectric}).

\section{Quantum mechanics with external potentials}

In 1990, Dyson published a demonstration of the Maxwell equations due
 to Richard Feynman \cite{dyson}. The demonstration
 dates back to the forties and had
 remained hitherto unpublished. It
 was believed to be incomplete because
 Feynman considered only the homogeneous Maxwell equations,
 given by the first two lines of Eq. (\ref{maxwell}):
\[
\bfn\ti\bE=-\pd_t\bB,\qquad \bfn\cdot\bB=0.
\]
 During the nineties, some authors revisited this demonstration and
 noted that the Schr\"odinger equation admitted external potentials only if
 they were compatible with the {\it magnetic} limit of LBLL
 and, therefore, with the Coulomb gauge condition (see
 \cite{brownholland,vaidyafarina} and references therein).

Indeed, from Eq. (\ref{maxmag}), it is clear that the homogeneous Maxwell
 equations given above are valid only within the magnetic limit
 because the electric field has zero curl in the electric limit, Eq.
 (\ref{maxel}). This is a consequence of the Galilean magnetic limit of the
 four-potential which does enter into the Schr\"odinger equation.
 Let us recall the statement more precisely (more details can be found in
 reference \cite{brownholland}).
 The Schr\"odinger equation with external fields $V(\bx,t)$ and
 $\bA(\bx,t)$ is written as
\[
\ri\hbar\partial_t\Psi(\bx,t)=
\frac 1{2m}(-\ri\hbar\nabla-\bA(\bx,t))^2\Psi(\bx,t)+V(\bx,t)\Psi(\bx,t).
\]
It is covariant under Galilean transformations, Eq. (\ref{xtgalilei}), with
 \[
\begin{array}{l}
{\Psi(\bx,t)\rightarrow \Psi'(\bx',t')}={\rm const}\; \exp[(\ri/\hbar)
 (-m\bv\cdot\bx+\frac 12m\bv^2t+\phi(\bx,t))]\; \Psi(\bx,t),\\
V(\bx,t)\rightarrow V'(\bx',t')=V(\bx,t)-\partial_t\phi(\bx,t)-\bv\cdot
 (\bA(\bx,t)+\nabla\phi(\bx,t)),\\
\bA(\bx,t)\rightarrow \bA'(\bx',t')=\bA(\bx,t)+\nabla\phi(\bx,t),\end{array}
\]
where $\phi(\bx,t)$ is some scalar function. In the case $\phi(\bx,t)=0$,
 which corresponds to pure Galilean boosts, the equations above reduce to
 the magnetic limit of Galilean transformations of the potentials,
 Eq. (\ref{magAVgalilei}). Hence, one can say
 that `Galilean covariance selects the gauge'.

 In a subsequent study, Holland and Brown have shown that the Maxwell
 equations admit an electric limit only if
 the source is a Dirac current \cite{brownholland2}. In addition,
 they have shown that the Dirac equation admits
 both Galilean limits, just like the Maxwell equations,
 corroborating thereby earlier
 results by L\'evy-Leblond \cite{levyleblond1967}. To summarize, what
 Feynmann did not (actually, could not) realize is that he had derived only
 the part of the Maxwell equations compatible with the Galilean covariant
 magnetic limit, that is to say, the homogeneous equations.

\section{Superconductivity}

Superconductivity also enters into the realm of the magnetic limit; indeed,
 it selects the Coulomb gauge condition as a necessary consequence of
 Galilean covariance. To illustrate this, consider the London equation,
 which states that the current density is proportional to the vector potential:
 \[
{\bf p }=m^* {\bf v}+q^* {\bf A}={\bf 0}.
\]
The star denotes a quantity describing Cooper pairs \cite{PGG}.
 This implies that there is a perfect transfer of electromagnetic momentum
 to kinetic momentum. Hence, contrary to what
 is usually stated, gauge invariance is not broken by superconductivity since the
 Coulomb gauge condition is implied. Moreover, the Meissner effect
 can be explained by starting with Amp\`ere's equation written as
 $\nabla\times\bB=\mu_0 \bj$, that is, without the displacement current term as
 in the magnetic case, third line of Eq. (\ref{maxmag}). Hence, this expression
 (or more directly $\nabla^2 {\bf A}\simeq -\mu_0 {\bf j}$ in the Riemann-Lorenz
 formulation) together with $\nabla\cdot\bA=0$ and London equation, lead to solutions
 (in one dimension $x$) of the type $A\simeq\exp{-{\lambda} x}$ (where $\lambda$ is
 a constant) so that the vector potential (hence the magnetic field) only penetrates
 the superconductor to a depth $1/{\lambda}$ \cite{PGG}.

We point out that the current density in the magnetic limit (hence in superconductivity)
 is divergenceless: by taking the divergence of
 \[
\nabla^2 {\bf A}\simeq -\mu_0 {\bf j}
\]
and using $\nabla\cdot\bA=0$, dictated by Galilean covariance,
 we end up with $\nabla\cdot\bj=0$. It does not mean, as often assumed, that the current
 is constant in time. Indeed, only the time derivative of the charge density is negligible
 with respect to the divergence of the current \cite{ejp}.

 As a consequence, superconductivity cannot be associated with a symmetry breaking
 of gauge invariance but is magnetic Galilean covariant. This unusual statement has been recently
 advocated by Martin Greiter using a different approach \cite{greiter}. As a matter
 of fact, it is the global U(1) phase rotation symmetry that is spontaneously
 violated. A striking consequence is that the Higgs mechanism for providing mass to particles becomes doubtful,
 since it was believed to be analogous to the assumed symmetry
 breaking of gauge invariance in superconductivity.

\section{Electrodynamics of continuous media at low velocities}

 In 1904, Lorentz claimed that a moving magnet could become electrically
 polarized \cite{Lorentz}. In 1908, Einstein and Laub noted that the
 Minkowski transformations for the fields and the excitations \cite{Minkowski} predict
 that a moving magnetic dipole induces an electric dipole moment \cite{EL}.
 It is interesting to reexamine these predictions in the light of the
 Galilean electrodynamics of continuous media. Indeed if one starts
 from the Minkowski transformations that relate the polarization and the
 magnetization \cite{Minkowski}, one would expect two Galilean limits:
 one with ${\bf{M}}' = {\bf{M}}$ and ${\bf{P}}' = {\bf{P}}-{\bf{v}}\times {\bf{M}}/c^2$
 and the other with ${\bf{M}}' = {\bf{M}}+{\bf{v}} \times {\bf{P}}$
 and ${\bf{P}}' = {\bf{P}}$ (see Chapter 9 of Ref. \cite{melba}).

In reference \cite{ejp}, we derived the following fields transformations:\\
\begin{table}[h!]
\begin{center}
\begin{tabular}{ccc}
\underline{Magnetic Limit} & \hspace{1in} & \underline{Electric Limit}\\
${\bf{B}} = {\bf{B}}'$ & & ${\bf{E}} = {\bf{E}}'$\\
 & & $\rho  = \rho '$\\
${\bf{j}} = {\bf{j}}'$ & & ${\bf{j}} = {\bf{j}}' + \rho '{\bf{v}}$\\
${\bf{H}} = {\bf{H}}'$ & & ${\bf{H}} = {\bf{H}}' + {\bf{v}} \times {\bf{D}}'$\\
${\bf{E}} = {\bf{E}}' - {\bf{v}} \times {\bf{B}}'$ & & ${\bf{D}} = {\bf{D}}'$\\
${\bf{M}} = {\bf{M}}'$ &  & ${\bf{P}} = {\bf{P}}'$\\
${\bf{P}} = {\bf{P}}'+{\bf{v}}\times {\bf{M}}'/c^2$ & & ${\bf{M}} = {\bf{M}}'-{\bf{v}} \times {\bf{P}}'$
\end{tabular}
\end{center}
\label{default}
\end{table}\\

We can infer the following boundary conditions for moving media,
 with ${\bf{n}}$ being the unit vector between two media denoted by 1 and 2,
 $\bf{K}$ the surface current, $\sigma$ the surface charge, $\Sigma$ the
 surface separating both media, and $v_n$ the projection of the relative
 velocity on the normal of $\Sigma$:
\begin{table}[h!]
\begin{center}
\begin{tabular}{ccc}
\underline{Magnetic Limit} & \hspace{1cm} & \underline{Electric Limit}\\
${\bf{n}} \times ({\bf{H}}_2  - {\bf{H}}_1 ) = {\bf{K}}$ & & ${\bf{n}} \times ({\bf{E}}_2  - {\bf{E}}_1 ) = 0$\\
${\bf{n}}\cdot({\bf{B}}_2  - {\bf{B}}_1 ) = {\bf{0}}$ & & ${\bf{n}}\cdot({\bf{D}}_2  - {\bf{D}}_1 ) = \sigma $\\
${\bf{n}}\cdot({\bf{j}}_2  - {\bf{j}}_1 ) + \nabla _\Sigma  \cdot{\bf{K}} = {\bf{0}}$ & &
 ${\bf{n}}\cdot({\bf{j}}_2  - {\bf{j}}_1 ) + \nabla _\Sigma
  \cdot{\bf{K}} = v_n (\rho _2  - \rho _1 ) - \partial _t \sigma $\\
${\bf{n}} \times ({\bf{E}}_2  - {\bf{E}}_1 ) = v_n ({\bf{B}}_2  - {\bf{B}}_1 )$ & &
 ${\bf{n}} \times ({\bf{H}}_2  - {\bf{H}}_1 ) = {\bf{K}} + v_n {\bf{n}}
 \times [{\bf{n}} \times ({\bf{D}}_2  - {\bf{D}}_1 )]$
\end{tabular}
\end{center}
\label{default}
\end{table}\\
Therefore, as we presumed, the effects in continuous media predicted by Lorentz as
 by Einstein and Laub are not purely relativistic since they can be
 described in a Galilean framework.

%%%%%%%%%%%%%%%%%%%%%%%%%%%%%%%%%%%%%%%%%%%%

%%%%%%%%%%%%%%%%%%%%%%%% ED of moving bodies @ low velocitites

%%%%%%%%%%%%%%%%%%%%%%%%%%%%%%%%%%%%%%%%%%%%

\section{Electrodynamics of moving bodies at low velocities}

Galilean electromagnetism raises doubts about our current understanding of
 the electrodynamics of moving media. For instance, several experiments
 (like the ones by Roentgen \cite{Roentgen}, Eichenwald \cite{Eichenwald},
 Wilson \cite{Wilson}, Wilson and Wilson \cite{WW}, Trouton and Noble
 \cite{trouton}, etc.) are generally believed to corroborate special
 relativity. However, as we will show hereafter for the Trouton-Noble experiment,
 there is not always a need for special relativity because the typical relative
 velocity in these experiments is much smaller than the speed of light.  As we
 have emphasized previously, the Galilean framework must involve
 the two limits of electromagnetism. A question that arises is
 which of the experiments mentioned above can be explained by either
 the electric limit, the magnetic limit or a coherent combination of both.

It is interesting to notice that Ernest Carvallo, a notorious anti-relativist,
 used both quasistatic limits as early as 1921 in order to deny the success of
 Einstein's theory of relativity \cite{Carvallo}. In some sense, he was right
 to point out that the electrodynamics of moving bodies at low velocities could
 be described in a Galilean-covariant manner by distinguishing the conductors
 and dielectrics. However, he was wrong to think that the optical properties of
 moving bodies can be described along the same lines.

%%%%%%%%%%%%%%%%%%%%%%%%%%%%%%%%%%%%%%%%%%%%%%%%%%%%%%%%

%%%%%%%%%%%%%%%% Trouton - Noble

\subsection{The Trouton-Noble experiment}

The Trouton-Noble experiment can be thought of as the electromagnetic
 analogue of the optical Michelson and Morley experiment \cite{trouton}.
 It was designed to verify whether one can observe a mechanical velocity of the
 ether if one considers the luminiferous medium as having parts which
 can be followed mechanically. Like the Michelson-Morley optical experiment,
 the Trouton-Noble experiment led to a negative results in the sense
 that no one was able to detect either an absolute motion with respect to the ether,
 or a partial entrainment like in Fizeau experiment.

In 1905, Albert Einstein suggested to consider the ether as superfluous, since its
 mechanical motion was not detected experimentally. Some theorists, such as
 H. Poincar\'{e} and H.A. Lorentz, were reluctant to relinquish ether as the
 bearer of the electromagnetic field, despite the fact that they had adopted the
 relativity principle. In 1920, at a conference in Leyden, Einstein himself recoursed
 to ether as the medium allowing the propagation of gravitational waves, although
 it cannot be endowed with the characteristics of a material medium \cite{leyden}.
 Today, even though the ether is a banished word in modern science, one can use it
 as did the older Einstein in order to describe the vacuum with physical
 (though not mechanical) properties.

Before the advent of special relativity, Hertz, Wien, Abraham, Lorentz,
 Cohn and all the specialists of the electrodynamics of moving bodies have
 used the transformations given in Eq. (\ref{wrongfieldtrans}), which is an
 incoherent mixture of the electric and magnetic Galilean limits \cite{darrigol2}.
 As mentioned previously, these expressions do not even obey the group property of
 composition of transformations.

The purpose of the Trouton-Noble experiment was to observe the effect
 of a charged capacitor in motion with an angle $\theta$ between
 the plates and the motion through the ether \cite{trouton}.
 The electric field in the reference frame of the capacitor
 generates a magnetic field in the ether frame given by
\[
\bB'=-\frac 1{c^2}\bv\ti\bE,
\]
where $\bv$ is the absolute velocity. Thus we have
\[
B'=\frac 1{c^2}vE\sin\theta.
\]
Consequently, there is a localization of magnetic energy density inside a volume d${V}$:
\[
{\rm d}W=\frac 12\frac{B'}{\mu_0}{\rm d}{ V}= \frac 12\frac{v^2}{c^2}\epsilon_0E^2
\sin^2\theta\ {\rm d}{ V}.
\]
The volume of the capacitor being written as $Sl$, the total energy between the plates is
\[
W=\frac 12\frac{v^2}{c^2}\epsilon_0E^2\sin^2\theta\ Sl.
\]
If one denotes by $V=E/l$ the difference of potential between the plates, then the
 capacitor is submitted to the electrical torque
\[
\Gamma=-\frac{{\rm d}W}{{\rm d}\theta}=-\frac{\epsilon_0}2\frac{V^2S}{l}
\frac{v^2}{c^2}\sin(2\theta),
\]
which is maximal for $\theta=45^\circ$, and zero for $\theta=90^\circ$.
 Hence, the plates are expected to be perpendicular to the velocity.
 However, this effect has not been observed experimentally.

In order to understand what is wrong with the above demonstration, let us first consider
 the electric limit transformation, given in Eq. (\ref{fieldselectric}). A consequence
 of this transformation is that the Biot-Savart law follows from the Coulomb law
 associated with the electric transformation of the magnetic field.
 Contrary to the transformations (\ref{wrongfieldtrans}), used by Trouton and Noble \cite{trouton},
 they do respect the group additivity. Besides, these transformations are only
 compatible with the approximate set of the Maxwell equations where the time derivative
 in the Faraday  equation vanishes, as in Eq. (\ref{maxel}).

We can derive the following `electric limit' approximation of the Poynting theorem:
\beq
\partial_t\left (\frac 12\epsilon_0E^2\right )+
\bfn\cdot\left (\frac{\bE\ti\bB}{\mu_0}\right )\simeq -\bj\cdot\bE.
\label{poyntingel}\eeq
This shows that the energy density is of electric origin only. Hence, no electric energy
 associated with the motional magnetic field can be taken into account within the
 electric limit, because it is of order $(v/c)^2$ with respect to the static, or
 quasistatic, electric one. Thus, the Trouton-Noble experiment does not show any effect
 as soon as we are in the realm of the electric limit. Recall that the electric limit
 is such that the relative velocity is small compared to the velocity of light $c$, and the
 order of magnitude of the electric field is large compared to the product of $c$ by
 the magnetic field. Of course, for larger velocities, special relativity is needed and
 we must take into account the additional mechanical torque \cite{PauliRel} due to the length variation
 in order to explain the negative result (i.e. no torque).

In the very last paragraph of their 1903 article, Trouton and Noble comment
 on the source of the negative result being caused by the fact that they
 considered the energy of the motional magnetic field \cite{trouton}.
 They suggested that the energy of the magnetic field must have had some origin,
 and that the electrostatic energy of the capacitor had to dimininish by
 $1/2 \epsilon _0 E^2 v^2/c^2$ when it is moving with a velocity $v$ at right angles
 to its electrostatic lines of force (the electrostatic energy then being
 $1/2 \epsilon _0 E^2$). We may assert that the converse situation of a solenoid or
 magnet in motion will not create a motional magnetic torque because the
 magnetic energy associated with the motional electric field is negligible compared
 to the magnetic energy of the static, or quasistatic, magnetic field.

 \subsection{`Einstein's asymmetry'}

In his famous article on the electrodynamics of moving media, Albert Einstein pointed out
 the importance of whether or not one should ascribe energy to the fields when
 dealing with motion \cite{Einstein}.
 In the introduction of his paper, he recalled that Maxwell's electrodynamics, when applied
 to moving bodies, leads to intrinsic theoretical asymmetries. He illustrated it with the
 example of the reciprocal electrodynamic action of a magnet and a conductor.
 Then, the observable phenomenon depends on the relative motion of the conductor
 and the magnet, unlike the traditional view advocated by Lorentz in which either one or the other of
 these bodies is in motion: (1) if the magnet is moving with the conductor at rest,
 an electric field, with a certain definite energy, is induced in the neighbourhood
 of the magnet, producing a current where parts of the conductor are located;
 (2) if the conductor is in motion and the magnet at rest, then no electric field
 arises in the neighbourhood of the magnet. Lorentz therefore argued that
 the conductor must contain an electromotive force with no intrinsic energy,
 but which causes electric currents similar to those produced by the electric
 forces in the former case, assuming the same relative motion in the two cases.
 This dual representation of the same phenomena was unbearable for Einstein.

By invoking the Lorentz transformation (obtained in the
 kinematical analysis of his article) to the Maxwell equations
 (actually, Einstein used the Heaviside-Hertz formulation,
 whereas Poincar\'{e} used the Riemann-Lorenz formulation in his
 relativity article \cite{Pierseaux}), Einstein replaced Lorentz's explanation: \\

{\em 1. If a unit electric point charge is in motion in an electromagnetic field,
there acts upon it, in addition to the electric force, an –electromotive force"
which, if we neglect the terms multiplied by the second and higher powers of
v/c, is equal to the vector product of the velocity of the charge and the magnetic
force, divided by the velocity of light.}\cite{Einstein}\\

\noindent{by the now famous special relativity explanation, valid for all velocities:}\\

{\em 2. If a unit electric point charge is in motion in an electromagnetic field,
the force acting upon it is equal to the electric force which is present at the
locality of the charge, and which we ascertain by transformation of the field to
a system of co-ordinates at rest relatively to the electrical charge.}\cite{Einstein}\\

Einstein therefore concluded that the analogy is valid with magnetomotive forces, based on
 the idea that the electromotive force is merely some auxiliary concept owing its
 existence to the fact that the electric and magnetic forces are related to the
 relative motion of the coordinate system. Then he pointed out that
 the asymmetry mentioned in the introduction of his article now disappears.

We wish to point out forcefully that the transformations of the electromagnetic field
 given by the Galilean magnetic limit are sufficient to explain Einstein's thought
 experiment with the magnet and the conductor, without recourse to Lorentz
 covariance \cite{norton}. Indeed, as in our discussion of the Trouton-Noble experiment,
 the magnetic Poynting theorem can explain why
 one cannot ascribe an energy to the motional electric field
 in Einstein's thought experiment,
\[
\partial_t\left (\frac{B^2}{2\mu _0}\right )+
\bfn\cdot\left (\frac{\bE\ti\bB}{\mu_0}\right )\simeq -\bj\cdot\bE,
\]
which is the magnetic analogue of Eq. (\ref{poyntingel}).

 It means that the second postulate (invariance of the velocity of light)
 used by Einstein is not required in order to explain the thought experiment. The relativity
 principle and the magnetic Galilean transformations are sufficient, together with the fact
 that the relative velocity involved in such an experiment is much smaller than the velocity
 of light. Hence, in the low-velocity regime, we proposed the following explanation of
 Einstein's asymmetry:\\

{\em 3. If a unit electric point charge is in motion in an electromagnetic field,
the force acting upon it is equal to the electric force which is present at the
locality of the charge, and which we ascertain by a Galilean magnetic transformation of the field to
a system of co-ordinates at rest relatively to the electrical charge.}\\

Einstein was correct in replacing Lorentz's explanation because Lorentz thought
 that the vector product of the velocity with the magnetic field was not an
 electric field (which is why Lorentz called it the electromotive field).
 But, because LBLL's article was not available yet, Einstein could not notice that
 the same vector product was an effective electric field due to a transformation of
 the Galilean magnetic limit. Wolfgang Pauli too offered a solution to the asymmetry
 problem in his textbook on electrodynamics, but he only assumed that his
 calculations were a first order approximation of the relativistic demonstration \cite{Pauli}.
 He did not acknowledge the existence of the Galilean magnetic limit.

To summarize, Einstein's explanation to remove the asymmetry is completely valid.
 However, we have noticed above that special relativity is not necessary to remove
 it, but only sufficient. It is ironic that the thought experiment that led
 Einstein to special relativity could have been explained by Galilean
 relativity if only the magnetic limit had been known by him at that time.

 As pointed out by Keswani and Kilminster \cite{Keswani}, Maxwell did resolve
 Einstein's asymmetry within the formalism of the magnetic limit when he stated that
 for all phenomena related to closed circuits and the current within them, the
 fact that the coordinate system be at rest or not is immaterial. On p. 346 of the
 same paper, they go on by explaining that the formula for the
 electromotive intensity (in its modern sense and not in Lorentz sense above) is
 of the same type, whether the motion of the conductors refers to fixed axes or to
 moving axes, because the only differences is that for moving axes the electric
 potential $V$ becomes $V'=V-{\bf v}\cdot{\bf A}$. Then
 they recall that Maxwell claimed that whenever a current is produced within a circuit $C$, the
 `electromotive force' is equal to $\int _C {\bf E'}\cdot d{\bf s}$, and the value of
 $V$ therefore disappears from this integral, so that the term $-{\bf v}\cdot {\bf A}$
 has no influence on its value \cite{Keswani}.

%%%%%%%%%%%%%%%%%%%%%%%%%%%%%%%%%%%%%%%%%%%%%%%%%%%%%%%%%%%%%%%%%%%%%%%%

%%%%%%%%%%%%%%%%%%%%%%%%%%%%       conclusion

\section*{Concluding remarks}

One century after the relativity revolution has taken place,
 and more than thirty years after the work of L\'evy-Leblond and Le Bellac,
 Galilean electromagnetism is becoming a field of current research,
 because it allows physicists and engineers to explain much
 more simply low-energy experiments involving the electrodynamics
 of moving media without the sophisticated formalism of special relativity.

In this paper, we have reexamined gauge conditions in connection with
 Lorentz and Galilean covariance. After a brief comment on the two Galilean
 limits of electromagnetism and the Faraday tensor, we have recalled the
 importance of the magnetic limit in `Feynman's proof'
 of the Maxwell equations as well as in superconductivity. Finally, we have questioned
 our current understanding of the electrodynamics of moving bodies by examining the
 Trouton-Noble experiment and the example used by Einstein in the introduction
 of his famous article on special relativity.

For slow velocities it is clear that effects of special relativity,
 such as length contraction, cannot explain (as it was believed so far)
 the corresponding experiments since these effects are negligible.
 In the realm of mechanics, one might ask what would have happened if Newton
 were born after Einstein? We are in an analogous situation
 with respect to electromagnetism.

\section*{Acknowledgement}

\noindent
M. de M. is grateful to NSERC (Canada) for financial support. G.R. was financially supported
 by a CNRS postdoctoral grant (S.P.M. section 02, France) during his stay in Nice.

%%%%%%%%%%%%%%%%%%%%  REFERENCES   %%%%%%%%%%%%%%%%%%%%%%%%%%%%%%%%%%%

\end{document}